\documentclass[fleqn,usenatbib]{mnras}

\usepackage{newtxtext,newtxmath}

\usepackage[T1]{fontenc}

\DeclareRobustCommand{\VAN}[3]{#2}
\let\VANthebibliography\thebibliography
\def\thebibliography{\DeclareRobustCommand{\VAN}[3]{##3}\VANthebibliography}

\usepackage{graphicx}	
\usepackage{amsmath}	

\title[Interlopers during the Cosmic Dawn]{On the expected purity of photometric galaxy surveys targeting the Cosmic Dawn}

\author[S.~R. Furlanetto \& J.~Mirocha]{
Steven R. Furlanetto$^{1}$\thanks{E-mail: sfurlane@astro.ucla.edu (SRF)}
\& Jordan~Mirocha$^{2}$
\\
$^{1}$Department of Physics and Astronomy, University of California, Los Angeles, CA 90095, USA \\
$^{2}$McGill University Department of Physics \& McGill Space Institute, 3600 Rue University, Montr\'eal, QC, H3A 2T8 \\
}

\date{Accepted XXX. Received YYY; in original form ZZZ}

\pubyear{2022}

\begin{document}
\label{firstpage}
\pagerange{\pageref{firstpage}--\pageref{lastpage}}
\maketitle

\begin{abstract}
Over the last three decades, photometric galaxy selection using the Lyman-break technique has transformed our understanding of the high-$z$ Universe, providing large samples of galaxies at $3 \la z \la 8$ with relatively small contamination. With the advent of the \emph{James Webb Space Telescope}, the Lyman-break technique has now been extended to $z \sim 17$. However, the purity of the resulting samples has not been tested. Here we use a simple model, built on the robust foundation of the dark matter halo mass function, to show that the expected level of contamination rises dramatically at $z \ga 10$, especially for luminous galaxies, placing stringent requirements on the selection process. The most luminous sources at $z \ga 12$ are likely at least ten thousand times rarer than potential contaminants, so  extensive spectroscopic followup campaigns may be required to identify a small number of target sources.
\end{abstract}




\section{Introduction}
\label{sec:intro}

In the past three decades, one of the most successful strategies for identifying distant galaxies has been photometric selection. The prototypical such method is the Lyman-break technique, which finds sources whose light blueward of the Lyman limit has been absorbed by the intergalactic medium (IGM), while light redward of that feature survives to the observer. Thus these galaxies appear luminous in filters redward of the Lyman limit but ``drop out'' of bluer filters; the ``break'' between these regimes then provides an estimate of the source's redshift \citep{Steidel1996}.  (At $z \ga 4$, the IGM is sufficiently neutral to absorb all light blueward of the Lyman-$\alpha$ transition rather than the Lyman limit, but by accounting for this difference the dropout technique can be extended to very high redshifts.) 

From the initial development of this method, it was recognized that, while photometric selection can identify large numbers of the target population of galaxies, it also has the potential for contamination. For example, with the Lyman-break technique, sources that are very dusty and red, but at lower redshift, can ``leak" through the selection and contaminate the sample. Fortunately, spectroscopic followup can measure the purity of the candidate population and improve the selection criteria; the most widely-used photometric techniques generally have only low levels of contamination. 

The ease with which photometric selection can be implemented has made it a valuable technique, one that has been modified to target a range of galaxy types at moderate and high redshifts. Very recently, it has been applied in a new regime -- at $z \ga 10$ -- to the early release data from the \emph{James Webb Space Telescope} (JWST; \citealt{Yan2022, Naidu2022, Leethochawalit2022, Donnan2022, Castellano2022, Atek2022, Adams2022, Harikane2022}). These analyses have identified several galaxy candidates at $z \ga 12$--17, many of them surprisingly luminous in comparison to theoretical expectations \citep{Labbe2022, Donnan2022, Castellano2022, Naidu2022}. The existence of these candidates has already challenged theoretical models of early galaxy formation \citep{Mason2022,Ferrara2022,BoylanKolchin2022,Lovell2022,Mirocha2022}. 

While compelling, these candidates reside in a new part of parameter space, for which the purity of photometric selection (i.e., the fraction of candidates that are true galaxies in the target population) has not yet been tested. Partly for this reason, most of these studies have attempted to further validate their candidates through spectral energy distribution (SED) fitting codes, such as \textsc{EAZY} \citep{Brammer2008}, \textsc{BEAGLE} \citep{Chevallard2016}, and \textsc{Prospector} \citep{Johnson2021}. These codes compare the observed photometry of each source to a library of template spectra in order to estimate the credibility interval of the source redshift. 

SED-fitting codes have been extensively tested against known galaxy populations, but of course their application to these new sources poses a new challenge: the inferred redshifts are only sensible if the template set is comprehensive, including both examples that match the target galaxy population (potentially problematic for early galaxies; e.g., \citealt{Steinhardt2022}) and those that match any potential interlopers from lower redshifts that could ``leak'' through the selection process. The latter problem has recently been highlighted by \citet{Zavala2022} and \citet{Naidu2022b}, who showed that unusually dusty star-forming galaxies with strong emission lines at $z \sim 5$ can mimic $z \sim 17$ dropout galaxies, and by \citet{Glazebrook2022} and \citet{Rodighiero2022}, who show that colour selection can reveal a wide range of source types. 

While it is widely recognized that these high-$z$ samples require eventual spectroscopic confirmation, there is little intuition for how pure the samples may be. Previously, photometric galaxy selection has been very successful, with purity $\ga 80\%$ even at $z \sim 8$, but it is not clear if that success can be extrapolated to the new regime at very high redshifts. 

Here, we use simple arguments to try to develop some intuition for the scale of the challenge. As we will show below, the expected number density of target galaxies at $z \ga 10$ is far smaller than the number density of galaxy populations that  could produce potential interlopers. In such a regime, even very unusual interloper galaxies -- potentially not (yet) included in SED template libraries -- could outnumber the target population. This has the potential to change the selection process from one in which interlopers are a nuisance, so that photometric samples (with SED fitting validation) are accurate enough to draw quantitative conclusions without source-by-source confirmation, to one in which spectroscopic confirmation is essential to making even preliminary conclusions. 

We outline our general method in section~\ref{sec:method} and show some  results to develop intuition in section~\ref{sec:results}. We then specialize to a particular case (in which Balmer-break galaxies contaminate Lyman-break targets) in section~\ref{sec:example}. We discuss the implications of our results in section~\ref{sec:disc}.

Throughout this work, we take $\Omega_m = 0.30$, $\Omega_\Lambda=0.7$, $\Omega_b=0.049$, $h = 0.68$, $\sigma_8=0.81$, and $n=0.9665$, consistent with the measurements of \citet{Planck2020}. All magnitudes are in AB units. 

\section{A simple estimate for the interloper population}
\label{sec:method}

Let us assume that we are seeking a ``target" population with an expected number count $N_t$ . There may be interlopers in the sample; let us assume that, for a given survey, their parent population has an expected number count $N_i$. In order to draw conclusions from the resulting candidates (without source-by-source confirmation), we require the final sample to be dominated by the target population. We thus need the fraction of the interlopers that leak through the selection process, $f_i$, to satisfy

\begin{equation}
 f_i N_i < N_t.
\end{equation}

For surveys targeting very luminous (and hence presumably rare) galaxies at high redshifts, we expect the interloper parent population to consist of fairly luminous sources at lower redshifts, where massive objects are much more abundant -- or in other words, $N_i \gg N_t$. Thus $f_i$ must be very small for the sample to be useful without spectroscopic confirmation. 

We can estimate the requirements placed on $f_i$ using simple arguments based on the halo mass function. Let us suppose the targets correspond to galaxies with stellar mass $M_{*,t}$. These live in halos with total masses
\begin{equation}
  M_{h,t} = M_{*,t}/(f_* f_b)
\end{equation}
where $f_b = \Omega_b/\Omega_m = 0.16$ is the baryon mass fraction, and $f_*$ is the star formation efficiency of the galaxy. The latter is unknown, but let us set it to a fiducial value of 10\% and parameterize it as $f_* = 0.1 f_{*,-1}$.

Next, we assume that the interlopers have halo masses
\begin{equation}
  M_{h,i} = \mu_i M_{h,t}.   
\end{equation}
We will consider the range $\mu_i \sim 1$--100 here. We will estimate this factor for a particular class of interlopers in section~\ref{sec:example}, but for now we simply note that if the targets and interlopers are comparably luminous they likely have similar stellar masses (and hence halo masses).  We take $\mu_i=10$ as our fiducial value, because it seems reasonable on theoretical grounds that star formation may be somewhat more efficient at the highest redshifts (e.g., \citealt{Furlanetto2017}). Note that, because the halo abundance declines with mass, a larger value of $\mu_i$ will decrease the abundance of interlopers. 

To obtain their number density, we must also specify the redshift at which the interloper population lives. As a simple, concrete example, let us assume that the targets are selected using the Lyman break dropout technique (which at very high redshifts occurs at the Lyman-$\alpha$ wavelength, $\lambda_\alpha=1216$~\AA), while the interlopers sneak in because of their Balmer break (at $\lambda_B = 3 \lambda_\alpha$). Then 
\begin{equation}
   (1 + z_i) \lambda_B = (1 + z_t) \lambda_\alpha
   \label{eq:zizt}
\end{equation}
which has the solution $(1 + z_i) = (1 + z_t)/3$. Other interlopers are certainly possible, but our choice is also close to the expectations for galaxies that have weak continua but strong optical emission lines (as explored in \citealt{Zavala2022, Naidu2022b}). 

With the masses and redshifts of the targets and interlopers in hand, we can use the halo mass function (which we write as $n_h(m,z)$, the comoving number density of halos with masses in the range $m \pm dm/2$ at redshift $z$) to estimate the number counts,
\begin{equation}
N_t \approx M_{h,t} n_h(M_{h,t},z_t) \Delta V_t,
\end{equation}
where $\Delta V_t$ is the comoving volume across the survey in which target objects can be detected, with a similar expression for $N_i$. We use the \citet{Trac2015} mass function for the target population, as it has been tested against simulations at $z>6$, and the \citet{Sheth1999} mass function for the interlopers. (Using the latter mass function for both regimes changes our results by $\la 2$.)

The volume factors depend upon the selection criteria and the nature of the interlopers, but the ratio of the volumes for the two populations should not be far from unity, and we will ignore it in our initial estimates. To see why, let us write
\begin{equation}
\Delta V_t \sim \left[ \Delta \theta D_A(z_t) (1+z_t) \right]^2 (dr/dz) \Delta z_t,
\end{equation}
where $\Delta \theta$ is the angular radius of the survey, $D_A$ is the angular diameter distance, $dr/dz$ is the comoving line element, $\Delta z_t = \Delta \lambda/\lambda_\alpha$ is the redshift range over which the relevant spectral feature can be detected, and $\Delta \lambda$ is the (observed) bandwidth. Taking the ratio of this volume to the corresponding expression for the interlopers, we have
\begin{equation}
    {\Delta V_t \over \Delta V_i} \approx \sqrt{3} \left[ {D_A(z_t) \over D_A(z_i)} \right]^2.
\end{equation}
For $z \gg 1$, the second factor is a (slowly) decreasing function of redshift, so the resulting factor will be near unity. Because the details will depend on the particulars of the selection functions and interloper populations, we ignore this factor when presenting our approximate results (though we do included it in the detailed example of section~\ref{sec:example}).

\begin{figure*}
\begin{center}
\includegraphics[width=0.98\textwidth]{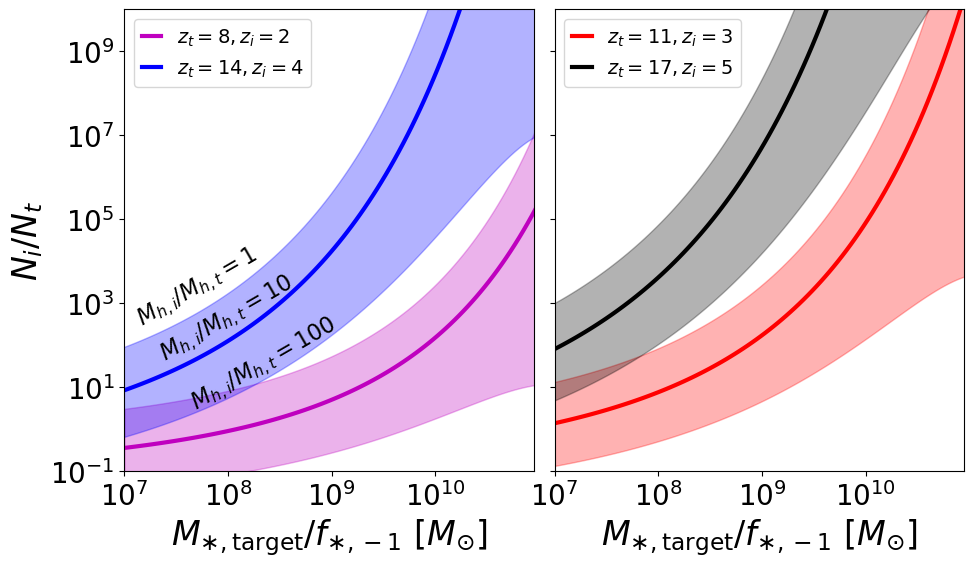}
\caption{{\bf Estimated requirements for interloper rejection in photometric surveys targeting the Cosmic Dawn.} We show the ratio of the counts of potential interlopers ($N_i$) to the counts of target galaxies ($N_t$); ideally, $N_t \gg N_i$. We consider surveys targeting $z=17, \, 14, \, 11$, and 8. In all cases, we choose the interloper redshifts by assuming that their Balmer break could be mistaken for a Lyman break. The curves assume that the interlopers reside in haloes ten times more massive than the target galaxies (or $\mu_i=10$); the shaded envelope shows the range for $\mu_i=1$--100. Note that we ignore the differential volumes between the two populations here, but that factor should be of order unity.}
\label{fig:interloper-fraction}
\end{center}
\end{figure*}

\section{Approximate requirements for candidate selection}
\label{sec:results}

Figure~\ref{fig:interloper-fraction} shows our main result: the ratio of the number of potential interlopers $N_i$ to the number of target galaxies $N_t$. With proper selection, this is of course far larger than the number of \emph{actual} contaminants. But the selection process needs to ensure $f_i N_i < N_t$ in order for a survey to be useful before spectroscopic confirmation. In other words, the fraction of interlopers that can ``leak" through the selection must be no larger than the reciprocal of $N_i/N_t$.


Across the two panels of Fig.~\ref{fig:interloper-fraction}, we show results for several redshifts ($z=8$--17) and for a range of halo mass ratios ($\mu_i$=1---100); the curves take $\mu_i=10$, with the shaded bands showing the full range. Note that the result is plotted as a function of the stellar mass of the target galaxy population, because that information is typically returned by SED-fitting codes (although their accuracy at high redshifts, where the rest-optical is poorly sampled, is questionable, especially if the stellar populations differ; \citealt{Steinhardt2022}). Because the abundance ratio actually depends on the halo in which this galaxy lives, this stellar mass must be transformed into a halo mass: to do so, we have normalized to our fiducial star formation efficiency ($f_{*,-1}=1$). 

Figure~\ref{fig:interloper-fraction} clearly shows that the potential for contamination increases dramatically through the Cosmic Dawn. While the requirements are quite modest at $z \sim 8$ -- as one would expect given the success of those surveys in the HST era -- our estimate suggests that the relative abundance of potential interlopers becomes orders of magnitude larger for massive galaxies at $z \ga 12$. The corresponding requirements on the care with which interlopers must be rejected become very stringent: if even one galaxy per ten thousand leaks through at $z_t \sim 14$ and $M_{*,t} \sim 10^9 \, M_\odot$, the sample will be dominated by interlopers. At $z \sim 17$ and $M_{*,t} \sim 10^9 \, M_\odot$, the acceptable leakage falls below one in a million! Even at $z \sim 11$, very massive galaxies ($M_{*,t} \ga 10^{10} \, M_\odot$) require exceptionally careful selection. That said, the requirements are much more modest for low-mass systems -- which remain relatively abundant at high redshifts. Interestingly, this suggests it may be more reliable to target faint candidates than bright ones!

This qualitative result is a direct consequence of the rapid evolution in the halo mass function during the Cosmic Dawn, especially at the massive end. While the precise form of the mass function at high redshifts is the matter of some controversy (e.g., \citealt{Reed2003,Trac2015, Mirocha2021}), this rapid evolution occurs in all similar calculations -- it is the key to hierarchical structure formation. 

That said, the requirement for $f_i$ is sensitive to the assumed mass ratio of the interloper and target halos, $\mu_i = M_{h,i}/M_{h,t}$. A larger mass ratio pushes the interlopers toward a less abundant set of haloes, so that a larger fraction can safely leak through. Understanding the properties of potential interloper populations, so that we can characterize their abundance, is important (see section~\ref{sec:example}). 

Our results also depend on the transformation from ``observed" stellar mass to halo mass. We have taken $f_*=0.1$ as a fiducial value, which is somewhat high compared to most low-redshift galaxies. A smaller star formation efficiency means that the target halo mass increases, making their population rarer (and worsening the contamination problem). But even assuming a maximal $f_*=1$, which effectively shifts all the curves to the right by one dex, still leaves extremely stringent requirements on the selection process for $z \ga 12$ (see also \citealt{BoylanKolchin2022}).

\begin{figure*}
\begin{center}
\includegraphics[width=0.48\textwidth]{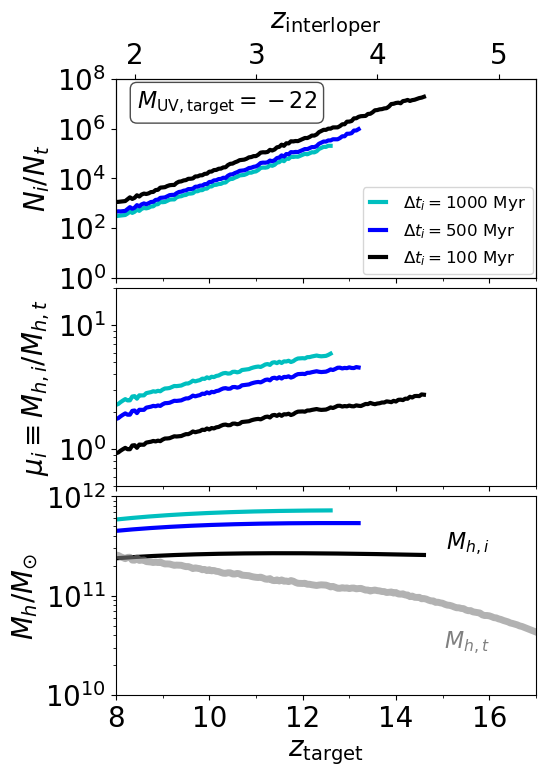}
\includegraphics[width=0.48\textwidth]{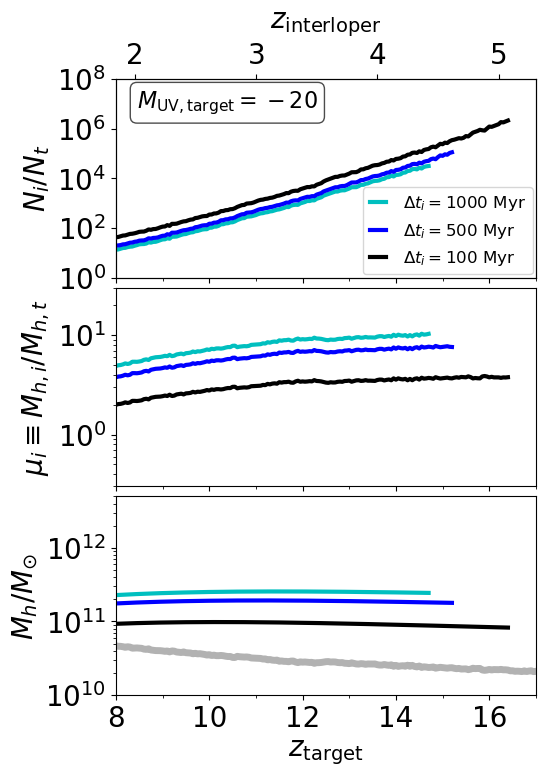}
\caption{{\bf Interloper abundance for high-$z$ target galaxies with a fixed absolute UV magnitude $M_{\rm{UV}}=-22$ (left) and $M_{\rm{UV}}=-20$.} The gray curves in the bottom panel show the inferred target halo masses from \textsc{ARES}; the other curves show the properties of the interlopers, with different choices for the age of the stellar population $\Delta t_i$. In all cases, the mass ratio between the interloper and target haloes is $\sim 1$--10, which results in the interloper parent population vastly outnumbering the target haloes, at least at these high luminosities.}
\label{fig:targetedMUV}
\end{center}
\end{figure*}

Given the early state of the JWST observations, it is difficult to know what a ``reasonable'' value of $f_i$ might be. The only spectroscopically confirmed galaxy at $z>10$ is GN-z11, which was discovered photometrically with the Hubble Space Telescope \citep{Oesch2014} and confirmed to be at $z \approx 11.1$ with a grism measurement of the continuum break at the location of the Lyman-$\alpha$ line \citep{Oesch2016}. This was the brightest of six objects identified in the photometric selection as laying at $z \sim 9$--12; the other candidates have not been confirmed to be at such high redshifts. The measurements also suggested $M_\star \sim 10^9 \ M_\odot$. According to Figure~\ref{fig:interloper-fraction}, $N_t > N_i$ only if $f_i \la 1$--10\% (assuming $\mu_i=10$), depending on the galaxy's $f_\star$. This suggests that SED template libraries are ``complete" to the percent level, which is reassuring. But the situation worsens rapidly at $z>12$ for galaxies of this mass scale.

We emphasize that our numerical results to this point should only be taken as guides to our intuition. Not only have we entirely ignored the details of the selection, we have also assumed that there is only a single population of potential interlopers. If we had allowed for \emph{any} galaxy to contaminate the samples, the requirements for $f_i$ would be even more stringent. Fortunately, in practice contamination of photometrically-selected samples seems to result from one or a small number of populations. Whether this remains true at the purity required for $z>12$ is an open question.

\section{An Example Interloper Population}
\label{sec:example}

The approach up to this point has left the target stellar mass and the relative masses of targets and interlopers as free parameters, in order to make the general point that the (potential) interloper population far exceeds the target population at $z \ga 12$. Of course, in practice a great deal is known about the properties of galaxies as a function of stellar mass across cosmic history, even for relatively distant potential interlopers. This knowledge can be leveraged to pin down their properties more precisely. In this section, we provide a more concrete worked example, in which we consider the possibility of confusion between a Lyman-break galaxy at very high redshift and a Balmer-break galaxy at moderate redshift. Our goal is to compute the expected number of potential interlopers for a target galaxy with a given UV magnitude $M_{\rm{UV}}$ at redshift $z_{t}$.

To do so, we take the following steps: 
\begin{itemize}
  \item First, we compute the redshift $z_i$ at which the interloper population must exist by comparing the Balmer breaks and Lyman break, as in the approximate calculation (see eq.~\ref{eq:zizt}).
  \item We then compute the luminosity required for the interloper to have the same apparent magnitude as the target in the observed band.
  \item Next, we assume that this corresponds to rest-frame 5000~\AA\ emission and compute the stellar mass required to provide this luminosity. For simplicity, we assume that this comes from a stellar population produced instantaneously in a burst a time $\Delta t$ before the interloper redshift. We consider $\Delta t_i= 100,\,500$, and 1000 Myr with \textsc{BPASS} version 1.0 single-star models \citep{Eldridge2009}. This allows us to compute Balmer breaks generated by galaxies with a range of star formation histories. 
  \item We then use results from the \textsc{UniverseMachine}  \citep{Behroozi2019}, a semi-empirical model calibrated to a slew of observations over cosmic history, which allows us to assign the required stellar mass to a corresponding halo mass at $z_i$. 
  \item Meanwhile, we also use \textsc{ares}\footnote{\url{https://github.com/mirochaj/ares}} \citep{Mirocha2020} to convert the chosen $M_{\rm{UV}}$ to a halo mass at $z_t$. This is necessary to estimate the intrinsic abundance of the sources and also allows us to consider other properties of the target population \citep{Mirocha2022}. 
\end{itemize}

Figure~\ref{fig:targetedMUV} shows the results of this procedure for two sets of galaxies: $M_{\rm UV}=-22$ (very luminous at these high redshifts) and $M_{\rm UV}=-20$ (moderately luminous). In both columns, the gray curve in the bottom panel shows the target halo mass needed to produce this luminosity. The other curves in the bottom panel show the inferred halo mass for the interloper population, for our three different assumed values of $\Delta t_i$. The curves are truncated when required stellar mass cannot be achieved at $z_i$ given the assumed stellar mass-halo mss relation from \textsc{UniverseMachine}; this point is of course sensitive to assumptions about the underlying stellar population, but even using the default parameters the point is clear. 

The middle panels show the inferred mass ratios $\mu_i$. Note that, across this set of target luminosities, $\mu_i \sim 1$--10 across the full redshift range, which motivates our fiducial choice in the previous sections. 

The top panels show the ratios of the interloper number densities to the target number densities for the different models. For very bright galaxies, it rises from $\sim 10^3$--$10^7$ from $z \sim 8$--15; for the fainter galaxies, it is 1--2 orders of magnitude smaller. The choice of $\Delta t_i$ has only a modest effect on the ratio, compared to the steep rise with redshift due to the underlying halo mass function. This occurs because galaxies with a larger $\Delta t_i$ must have formed \emph{more} stars to achieve the same brightness at $z_i$ than galaxies with a small $\Delta t_i$, because their more massive stars have already died; as a result, a large $\Delta t$ implies a large halo mass (where $f_{\ast}$ is large) and so a less abundant interloper population. These large ratios are qualitatively consistent with our estimates in Figure~\ref{fig:interloper-fraction}, though factoring in the evolution of the target and interloper populations does moderate the dependence on with redshift. 

Regardless, the overall conclusion remains the same: the population of potential interlopers vastly outnumbers the population of target haloes, increasing dramatically with redshift and target halo mass. The required level of ``cleaning" through SED fitting (for example) also increases, so that the library of template spectra for the interloper population must include not only ``typical" members of the class but also rare outliers in their population. 

\section{Discussion}
\label{sec:disc}

We have used a very simple model -- which essentially relies only on the evolution of the halo mass function -- to provide some intuition for the challenges to selecting very high-$z$ galaxies against a background of their ``low''-$z$ neighbors. Of course, spectroscopic confirmation will still reliably select the target population, but short of that we must ask how to interpret photometrically-selected candidates. Because the halo mass function evolves so rapidly during the Cosmic Dawn, especially at large masses, the problem of finding galaxies will inevitably become harder and harder at higher redshifts. By $z \sim 14$, the population of population interlopers will outnumber the haloes in which massive galaxies are expected to live by many orders of magnitude -- requiring searches to reject potential interlopers at extremely high efficiency. 

The stringent requirements for candidate rejection suggested by our model are concerning in (at least) two ways. First, if interlopers must be rejected at the level of (e.g.) 1 in $10^4$, the template library to which a comparison is done must include examples that deviate by more than ``$3\sigma$'' (however that might be precisely defined) from typical galaxies. In other words, it does not suffice to show that a contaminant could only be produced by a ``very rare'' galaxy if the ``very rare'' subset of the interloper population is just as numerous as the target population. The one spectroscopically confirmed $z>10$ galaxy, GN-z11 \citep{Oesch2016}, suggests that template libraries are complete to (at least) the percent level, which is heartening -- but still far below the requirements for massive galaxies at $z \ga 14$. 

The second immediate problem is the danger that the interlopers may be a class of rare but heretofore unrecognized sources -- examples of such populations are still being discovered at a relatively rapid pace at high redshifts (e.g., the ``HST-dark" population; \citealt{Barrufet2022}). Such galaxy populations are by definition missing from SED libraries, and even very exceptional groups pose challenges for galaxy selection at high redshifts. 

Of course, we are not entirely helpless in the face of this challenge. In section \ref{sec:example} we saw that existing models can help guide our understanding of potential interloper populations and so help determine (or even model) more detailed requirements on sample purity. One possible avenue will be to optimize  the construction of template-fitting codes to make redshift inferences more robust. For example, one must ensure that the corners of the hyperspace spanned by the template spectra parameters are explored fully enough to rule out rare interlopers. 

The challenge of finding rare galaxies in large samples emphasizes the importance of spectroscopic followup for identifying the earliest sources in the Universe. This does not by any means devalue the initial selection, but it suggests caution in interpreting those results (at least until the sample purity of early surveys can be measured). An analogy can be made to the selection of high-$z$ quasar candidates from large photometric surveys, for which the spectroscopic success rate is small even after more than two decades of refinement (e.g., \citealt{Wang2019}) -- but the sources that have been found have led to extraordinary insights, motivating large-scale, dedicated spectroscopic followup campaigns. 

The model presented here also has implications for designing these spectroscopic followup campaigns. Because the halo mass function evolves much more rapidly at high masses, our results suggest that the sources most likely to be confirmed successfully are of relatively modest mass and luminosity -- motivating prioritizing ``normal'' galaxies in the followup. Of course, by some metrics the most massive galaxies are the most exciting -- even though they are most subject to contamination. Surveys should balance the intrinsic interest of these sources with the likelihood of successful confirmation. 

Finally, it must be emphasized that the guidance advocated in this paper is of course only relevant if the underlying theoretical models are accurate. In particular, if the Universe has many more massive haloes than expected, our estimates will be wrong. This is already one potential interpretation of the JWST data, which will also be tested by future followup. 

\section*{Data Availability}

No new data were obtained as part of this work. Results used to generate the figures are available upon request from the authors.

\section*{Acknowledgments}

SRF was supported by the National Science Foundation through award AST-1812458. In addition, this work was directly supported by the NASA Solar System Exploration Research Virtual Institute cooperative agreement number 80ARC017M0006. This work has made extensive use of NASA’s Astrophysics Data System (http://ui.adsabs.harvard.edu/) and the arXiv e-Print service (http://arxiv.org).



\bibliographystyle{mnras}
\bibliography{interloper} 




\bsp	
\label{lastpage}
\end{document}